\documentstyle[preprint,aps]{revtex}
\begin{document}
 \setcounter{page}{1}
\title{Absence of Bose-Einstein condensation with a uniform field and contact interaction}
\author{Roberto Soldati$^1$ and Alberto Zannoni$^2$}
\address{$^1$ Dipartimento di Fisica ``A. Righi'', Universit\`a di Bologna and
Istituto Nazionale di Fisica Nucleare, Sezione di Bologna, 40126 Bologna - Italy}
 \address{$^2$IPST, University of Maryland, College Park, MD 20742,USA and
Electron and Optical Physics Division, NIST,Gaithersburg, MD 20899, USA}

\date{4-July-2000}
\maketitle

\begin{abstract}
The behavior of a Bose-Einstein ideal gas of particles in a three dimensional space
in the presence of a uniform field, such as gravity, and of contact interaction, describing the presence of
one impurity, is investigated. It is shown that Bose-Einstein condensation can not occur.
\end{abstract}
\pacs{PACS numbers: 03.75.F, 05.30}

The recent experiments \cite{Cornell} on Bose-Einstein Condensation (BEC) in
atomic vapor magnetically trapped, have stimulated a new interest in the
theoretical study of Bose gases \cite{sandro}.
In this Letter we propose a theoretical model,
that could be relevant to simulate what
happens in the presence of some particular external fields
such as a gravitational uniform
field and a point-like (or at ``zero
range'', or $\delta$-like) interaction, which describes the presence of impurities.
We start from the classical one-particle Hamiltonian of the problem, that is

\begin{equation}
H=\frac{ {\bf {p}}^2 }{2m}+ m g x_3\; ,
\label{a}
\end{equation}

\noindent where $m$ and ${\bf{p}}=(p_{1},p_{2},p_{3})$ are the mass and the
momentum of the particle respectively. The contact interaction, {\it i.e.} the
presence of the impurity, will  be included in the formalism after
quantization,
which can be done only under the specification of particular boundary
conditions on the wave function satisfying the Schr\"odinger equation
\cite{Albeverio}. After the introduction of dimensionless cylindrical coordinates,

\begin{equation}
x_1 = \frac{r}{\kappa}\cos{\phi}\;,\;
x_2 = \frac{r}{\kappa}\sin{\phi}\;,\;
x_3 = \frac{z}{\kappa}
\end{equation}
\begin{equation}
0<\phi<2\pi\;,\; 0<r <\infty \;,\; -\infty< z <\infty
\label{b}
\end{equation}

\noindent with the quantum gravitational length and energy defined by

\begin{equation}
\label{k}\lambda_g\equiv\kappa^{-1}=\left(\frac{\hbar^2}{2m^2 g}\right)^{1/3}\; ,
\qquad E_g=\frac{\hbar^2\kappa^2}{2m}\; ,
\end{equation}

\noindent the eigenvalues equation $H\Psi_E=E\Psi_E$ gives us the following
general form of the eigenfunctions in terms of Bessel's, Neumann's and Airy's functions \cite{abra}:

\begin{eqnarray}
&\Psi_{l,\lambda,\epsilon}(r,z,\phi)=Z(\eta)e^{il\phi}R_l(\lambda r)/\sqrt{2\pi}\; ,\\
\label{c}&\eta = z-\epsilon+\lambda^2\; ,\\
\label{wfr}&R_l(\lambda r)=A_{R,l}J_{l}(\lambda r)+B_{R,l}N_{l}(\lambda r)\; ,\\
\label{wfz}&Z(\eta)=A_{Z} Ai(\eta) + B_{Z} Bi(\eta)\; ,
\end{eqnarray}

\noindent where $A_{R,l},B_{R,l},A_Z,B_Z$ are normalization constants
to be further determined,
whilst the dimensionless quantum numbers are defined to be

\begin{equation}
\epsilon=\frac{E}{E_g}\in{\bf{R}}\ ,\qquad
l\in{\bf Z}\ ,\qquad
\lambda=\sqrt{\frac{p_x^2+p_y^2}{\hbar^2\kappa^2}}\ge 0\, .
\end{equation}

\noindent In order to evaluate the
normalization constants we have to impose the usual
orthonormality relations

\begin{eqnarray}
\label{z}
&\int_{-\infty}^{+\infty}dz
Z(z,\epsilon,\lambda)Z(z,\epsilon ',\lambda) =
\delta(\epsilon-\epsilon ')\\
&\int_{0}^{\infty}
dr\; rR_{l}(\lambda r)R_{l}(\lambda' r)=
\delta(\lambda - \lambda ') \label{d}
\end{eqnarray}

\noindent
Concerning eq.~(\ref{d}), it is necessary to separate the cases with
$l\neq0$ and $l=0$. This distinction is the main point of the discussion. As a matter of fact, we have
to carefully take into account that the particles are in a
uniform field and that they feel, from a quantum mechanical point of
view, another extremely localized interaction (point-like or
$\delta$-like interaction) usually called contact interaction describing the impurity. We know \cite{Albeverio} that
$\delta$-like potentials on the plane are not mathematically well defined in
quantum mechanics. The only correct quantum mechanical formalism to describe
them is in terms of the self-adjoint extensions of the
Hamiltonian operator which, in the present case, is given by the differential operator
of eq.~(\ref{a}). Actually, that differential operator is only symmetric, until
we specify its domain. These self-adjoint extensions of the
Hamiltonian are non-trivial in the case $l=0$, when both the regular and
the irregular wave functions at $r=0$ are allowed, in contrast with the case
$l\neq0$, in which singularities at $r=0$ can not be accepted.
It is indeed possible to see from the behavior of the Neumann's function for large argument
\cite{Grad} that $B_{R,l}=0$, for the wave-function has to be locally
square integrable. Therefore we readily get

\begin{equation}
A_{R,l}(\lambda)=-\sqrt{\lambda}\; ,\qquad l\neq 0\; .
\end{equation}

\noindent In the case $l=0$ the presence of the
irregular part of the wave function is allowed
since it turns out to be locally square integrable.
A straightforward calculation leads to the result that the normalization condition
(\ref{d})  is true if and only if

\begin{equation}
\label{alpha}\pi\tan[\pi\nu(\alpha_0,\lambda)]=\ln(\alpha_{0}/\lambda^2)\; ,
\qquad \lambda\ge 0\ ,
\end{equation}

\noindent where $0\le\alpha_{0}$ is an arbitrary constant parameter. Moreover,
taking eq.~(\ref{alpha}) into account we definitely obtain

\begin{eqnarray}
A_{R,0}(\lambda)=-\sqrt{\lambda}\sin(\pi\nu) \;,\; B_{R,0}(\lambda)=\sqrt{\lambda}\cos(\pi\nu)\; .
\end{eqnarray}

\noindent Now it is very important to understand the role of $\alpha_{0}$. As
we have already emphasized, we are dealing with a quantum-mechanical
point-like interaction,  which can be correctly handled through
the formalism of the self-adjoint extensions of the Hamiltonian operator
\cite{Albeverio}. The latter formalism entails that in order to describe the dynamics
a whole one-parameter continuous family of different quantum mechanical Hamiltonians
has to be considered. The different self-adjoint Hamiltonian operators are
naturally labelled by the parameter $\alpha_{0}$. When
$\alpha_{0}=0$ (Friedrichs' limit), the contact interaction is
turned off and, from the  mathematical point of view, it means that the
irregular part of the wave-function disappears. Actually

\begin{equation}
\tan(\pi\nu)=-\infty \Rightarrow\nu=-{1/2}\; ,
\end{equation}

\noindent {\it i.e.}, the same result as in the case $l\neq0$.
Let us finally evaluate the constants $A_{Z}$,$B_{Z}$.
Substituting eq.~(\ref{wfz}) into eq.~(\ref{z}),
and noting that from the asymptotic behavior of the Airy's function $Bi(z)$ we
have to set $B_{Z}=0$ - in order to keep square summability on the positive half-line $z>0$
- we finally  get

\begin{equation}
(A_{Z})^2 \int_{-\infty}^{\infty} dz Ai(z,\epsilon,\lambda)
Ai(z,\epsilon',\lambda)=\delta(\epsilon-\epsilon ')\; .
\end{equation}

\noindent Using the integral representation for the Airy's function \cite{Landau}
we find

\begin{equation}
A_{Z}=1/\sqrt{2\pi} \; , \; B_{Z}=0\; ,
\end{equation}

\noindent
the wave functions of the eigenstates with vanishing angular
momentum and with a continuous degeneracy $\lambda\ge 0$ due to the transverse momentum
becomes

\begin{equation}
\label{wf1}\Psi_{0,\lambda,\epsilon}(r,z,\phi)
=\frac{\sqrt{\lambda}Ai(\eta)}{2\pi \sec{(\pi\nu)}}\left[N_{0}(\lambda r)-\tan{(\pi\nu)}J_{0}(\lambda r)\right]\; ,
\end{equation}

\noindent whereas the eigenfunctions of the states with non-vanishing angular
momentum read

\begin{equation}
\label{wf1m}\Psi_{l,\lambda,\epsilon}(r,z,\phi)=-
\frac{\sqrt{\lambda}}{2\pi}J_{l}(\lambda r)Ai(\eta)e^{il\phi}\; .
\end{equation}

\noindent From the previously obtained values for the normalization constants, it follows
that the improper eigenfunctions are normalized according to

\begin{equation}
\label{on}\left<\Psi_{l,\lambda,\epsilon}|\Psi_{l^\prime,\lambda^\prime,\epsilon^\prime}\right>
=\delta_{l,l^\prime}\delta(\lambda-\lambda^\prime)\delta(\epsilon-\epsilon^\prime)\ .
\end{equation}

Now we are ready to discuss the existence of a characteristic
state which faithfully and uniquely specifies any given self-adjoint extension,
within the above mentioned one-parameter continuous family of the quantum
Hamiltonian operators that are allowed according to the general principles of
quantum mechanics.
To this purpose, let us start again  from the Schr\"odinger stationary
equation, after the replacement $\lambda\rightarrow i\hat\lambda$.
Then we obtain

\begin{eqnarray}
\label{nr}\frac{r}{R_l(r)}{\partial \over {\partial r}} \left(r {\partial
\over {\partial r}}\right)R_l(r)- \hat{\lambda}^2r^2-l^2 &=& 0\; ,\\
\label{nz}\frac{1}{Z(z)}{\partial^2 \over {\partial z}^2}Z(z)-z+
{\epsilon} + \hat{\lambda}^2 &=& 0\; .
\end{eqnarray}

\noindent The first equation is exactly the same that we had previously discussed
in the case of states with a continuous degeneracy in the transverse momentum.
Our interest relies only in the subspace of vanishing angular momentum because it is only in this subspace that non-trivial
self-adjoint extensions of the Hamiltonian actually occur. For $l\neq0$
the solution of eq.~(\ref{nr}) is not acceptable because it does not belong to
$L^2({\bf{R}}^3)$. Taking $l=0$ and evaluating the wave function of the
state that characterizes the extension we get

\begin{equation}
\label{wfce}\hat{\Psi}_{0,\epsilon}(r,z)=
\frac{\mid \hat{\lambda} \mid }{\sqrt{2\pi^2}}K_{0}(\mid
\hat{\lambda} \mid r)Ai(z-\hat{\epsilon} -\hat{\lambda}^2)\; .
\end{equation}

\noindent This state has to be orthogonal to any of the improper
states with a continuous degeneracy of the transverse
momentum $\lambda\ge 0$ - see eq.~(\ref{wf1}). We expect to find that the
orthogonality takes place when the parameter $\hat{\lambda}^2$ of the
state characterizing the self-adjoint extensions is exactly equal to the above
introduced parameter $\alpha_{0}$. Actually we require

\begin{equation}
\left<\Psi_{0,\lambda,\epsilon}|
\hat{\Psi}_{0,\epsilon^\prime}\right>=0\; ,\qquad
\forall \; \epsilon\neq\epsilon^\prime\ ,\quad \lambda\ge 0\; .
\end{equation}

\noindent Explicit evaluation leads to the result

\begin{equation}
\pi\tan(\pi\nu)=\ln(\hat{\lambda}^2/\lambda^2)\; ,
\end{equation}

\noindent which exactly corresponds to the definition of $\alpha_{0}$ in
eq.~(\ref{alpha}). Consequently,
orthogonality occurs iff $\hat{\lambda}^2=\alpha_{0}$, and from now on
we can label the state that uniquely characterizes the extension with
the parameter $\alpha_{0}$.\\

The calculation of the
one-particle Heat-Kernel is the first step to get
the partition function of the system. It is also interesting to examine
the limiting cases in which the background fields are switched off. As a
matter of fact, we have to recover the Heat-Kernel of the
non-relativistic free particle when both of the external fields are turned off,
{\it i.e.}, $g\rightarrow 0$ and $\alpha_{0}\rightarrow 0$ . To get the
general form of the Heat-Kernel, we need the spectral
decomposition of the quantum Hamiltonian operator.
To this purpose, it is necessary to keep in mind that we have
to deal both with the states with a continuous degeneracy, labelled by the
eigenvalues of the transverse momentum, and with the single state
characterizing the self-adjoint extension. Furthermore, it is convenient
from now on to measure energies in units of quantum gravitational energy scale
$E_g$.
As a consequence, we define the dimensionless Hamiltonian operators as

\begin{equation}
E_g^{-1}H(\alpha_0)\equiv
{\rm H}(\alpha_{0})=\check{\rm H}[\nu(\alpha_{0})]+\hat{\rm H}(\alpha_{0})\; ,
\end{equation}

\noindent where $\check{\rm H}$ is the part of the spectral
decomposition of the Hamiltonian that involves the states with a continuous
degeneracy of the transverse momentum, and $\hat{\rm H}$ is the one that involves
the characteristic state of the self-adjoint extension.
The explicit form of the spectral decompositions can be written as

\begin{eqnarray}
\label{hc}&\check{\rm H}[\nu (\alpha_{0})]=\int_{-\infty}^{+\infty}d\epsilon \;\epsilon \int_{0}^{\infty}d\lambda\sum_{l\in {\bf Z}}|\epsilon,\lambda,l><\epsilon,\lambda,l|\; ,\\
\label{hh}&\hat{\rm H}(\alpha_{0})= \int_{-\infty}^{+\infty}d\epsilon \; \epsilon \mid
\epsilon,\alpha_{0},0> <\epsilon,\alpha_{0},0 \mid\; .
\end{eqnarray}

\noindent
Let us come back to the evaluation of the diagonal kernel. For a matter of
simplicity,
we shall calculate separately the part of the kernel corresponding to the
continuous degeneracy in the transverse momentum, and the other one
which characterizes the extension: namely,

\begin{eqnarray}
G(\alpha_0,\gamma;{\bf r})&=&\check{G}(\alpha_0,\gamma;{\bf r})+\hat{G}(\alpha_0,\gamma;{\bf r})\ ,\\
\label{gc}\check{G}(\alpha_0,\gamma;{\bf r}) &=& e^{\gamma z}\left<{\bf r}\mid
\exp\{-\gamma \check{\rm H}[\nu(\alpha_0)]\}
\mid {\bf r}\right>\ ,\\
\label{gh}\hat{G}(\alpha_0,\gamma;{\bf r}) &=& e^{\gamma z}\left<{\bf r}\mid \exp\{-\gamma \hat{\rm H}(\alpha_0)\}\mid
{\bf r}\right>\ ,
\end{eqnarray}

\noindent where ${\bf r}=(r,\phi,z)$ is the position in the dimensionless coordinate space
associated to the system whereas
$\gamma=\beta E_g$ and $\beta\equiv 1/k_BT$,
with $k_{B}$ the Boltzmann's constant. The reason for the factor $\exp\{\gamma z\}$ in eq.s~(\ref{gc}-\ref{gh})
is that one has to recover translation invariance in the absence of contact interaction, {\it i.e.}, in the limit $\alpha_0\to 0$.
The diagonal kernel $\check{G}$ is that one with a continuous
degeneracy in the transverse momentum. From eq.s~(\ref{gc}) and (\ref{hc}) we
can write

\begin{equation}
\check{G}(\alpha_0,\gamma;{\bf r})= \int_{-\infty}^{\infty} d\epsilon \; e^{\gamma (z-\epsilon)}
\int_{0}^{\infty}d\lambda
\sum_{l\in{\bf Z}}\left|\Psi_{l,\lambda,\epsilon} (r,z,\phi)\right|^2
\label{gcheck}
\end{equation}

\noindent Substituting the wave functions (\ref{wf1})-(\ref{wf1m}) into the
expression (\ref{gcheck}) we obtain

\begin{eqnarray}
\nonumber\check{G}(\alpha_{0},\gamma;{\bf{r}})&=&
\frac{1}{(2\pi)^2}\int_{-\infty}^{\infty} d\eta \; e^{\gamma \eta} Ai^2(\eta)
\int_{0}^{\infty} d \lambda \lambda  e^{-\gamma \lambda^2}\\
&\times&\nonumber \left\{1-\cos^2(\pi\nu)\left[J^2_{0}(\lambda r)-N^2_{0}(\lambda r)\right]\right . \\
&+&\left . \sin(2\pi\nu)J_{0}(\lambda r)N_{0}(\lambda r)\right\} \, .
\end{eqnarray}

\noindent The integral over the Airy's function can be further elaborated and
eventually set in the more suitable form
\noindent where

\begin{equation}
\label{Airint}I(\gamma)\equiv \int_{-\infty}^{+\infty}d\eta \; e^{\gamma \eta} Ai^2(\eta)=I_{1}+I_{2}+I_{3}\; ,
\end{equation}

\noindent with
\begin{eqnarray}
\label{I1}I_{1}&=\frac{1}{18}\sqrt{\frac{3}{\pi}}\left(\frac{3}{2}\right)^{\frac{1}{3}}
E_{3}\left[\frac{\gamma}{3}\left(\frac{3}{2}\right)^{\frac{2}{3}},\frac{7}{6}\right]\; ,\\
\label{I2}I_{2}&=\frac{1}{\sqrt{2\pi}}\int_{-\infty}^{\infty}d\xi
\frac{\gamma}{\sqrt{\mid\xi\mid}(\gamma^2+\xi^2)}\cos{\frac{\xi^2}{12}}\; ,\\
\label{I3}I_{3}&=\frac{1}{\sqrt{2\pi}}\int_{-\infty}^{\infty}d\xi
\frac{(-\gamma)}{\sqrt{\mid\xi\mid}(\gamma^2+\xi^2)}\sin{\frac{\xi^2}{12}}\; ,
\end{eqnarray}
\noindent $E_{\rho}(z,\mu)$ being the Mittag Leffer function \cite{Grad}.

Finally, taking the above integral representation into account, we can rewrite
the Heat-Kernel $\check{G}$ in the simpler form

\begin{eqnarray}
&\nonumber \check{G}(\alpha_{0},\gamma;r) =I(\gamma)/(8\pi^2\gamma)
\left\{ 1-\int_0^\infty dt\ e^{-t} \cos^2(\pi\nu)\right . \\
&\nonumber  \left .\times \left[J^2_{0}\left(s\sqrt{t}\right)
-N^2_0\left(s\sqrt{t}\right)\right]
-\sin(2\pi\nu)J_{0}\left(s\sqrt{t}\right)
N_{0}\left(s\sqrt{t}\right)\right\}\; ,\\
&s=\frac{\sqrt{4\pi(x_1^2+x_2^2)}}{\lambda_T}\; , \;\pi\tan(\pi\nu)=\ln(\alpha_0\gamma/\ln 
t)\, , \end{eqnarray}
where $\lambda_{T}$ is the thermal wave length.

The evaluation of  the diagonal kernel $\hat{G}$
can be done in close analogy with
the previous case. From eq.~(\ref{wfce}) with $\hat{\lambda}^2$
replaced by $\alpha_{0}$ and from eq.s~(\ref{hh}) and (\ref{gh}) we get

\begin{equation}
\hat{G}(\alpha_{0},\gamma;r)=
\frac{\alpha_{0}}{2\pi^2} e^{\gamma\alpha_0}
K_{0}^2(r\sqrt{\alpha_{0}})I(\gamma)\; .
\end{equation}

\noindent
It is very interesting now to examine the behavior
of the kernel in the limiting cases in which the background fields are switched
off. Particularly, we look at the behavior of the diagonal Heat-Kernel when
$g\rightarrow0$, {\it i.e.}, when the uniform field is off, and when
$\alpha_{0}\rightarrow 0$, that means without point-like impurity. Finally we
will examine the case when both the external fields are turned off and we show
that there is commutativity with respect to order of the operations.
We do emphasize that, in order to evaluate those limits, it is
necessary to pass from the dimensionless Heat-Kernel, to the corresponding one
in physical units: namely

\begin{equation}
\label{Gtilde}
\tilde{G}(\alpha_0,g,T;{\bf x})=
\kappa^3 G(\alpha_0,\gamma; {\bf r})=\left(\frac{\sqrt{4\pi\gamma}}{\lambda_T}\right)^3
G(\alpha_0,\gamma; r)\; ,
\end{equation}

\noindent where the dimensional quantities are labeled with a tilde.
From the behavior for small $\gamma$, of the integral $I(\gamma)\sim\sqrt{(\pi/\gamma)}$
- see eq.s (\ref{Airint}-\ref{I3}) - it immediately follows that

\begin{equation}
\label{nuova}\tilde{G}(g\to 0) =\lambda_T^{-3}\; .
\end{equation}

\noindent which shows that the above discussed specific form of the contact interaction
does completely depend upon the presence of a non-vanishing uniform field.
In order to perform the limit in which the point-like impurity is
removed, it is important to keep in mind that $\alpha_{0}$ and $\nu$ are strictly
related by eq.~(\ref{alpha}), so that $\nu$ goes
to $(-1/2)$ when $\alpha_{0}$ goes to zero.
The explicit evaluation yields

\begin{equation}
\label{aoff}\tilde{G}(\alpha_{0}\rightarrow 0)=
\frac{I(\beta E_g)}{\lambda_T^3}\sqrt{\frac{\beta E_g}{\pi}}\; ,
\end{equation}
which exhibits manifest translation invariance as it does.
From eq.s~(\ref{nuova}) and (\ref{aoff}), after
switching off the remaining background fields we get the expected
result

\begin{equation}
\lim_{g\to 0}\tilde{G}(\alpha_0,g,T;{\bf x})=
\lambda_{T}^{-3}=
\lim_{g\to 0}\lim_{\alpha_0\to 0}\tilde{G}(\alpha_0,g,T;{\bf x})\ ,
\end{equation}

\noindent {\it i.e.}, the Heat-Kernel of a free non-relativistic particle.
Furthermore we have shown that the final result is independent from the order
with which the external fields are switched off.\\

The partition function {\it per} unit volume of a single molecule
is  now to be evaluated in the thermodynamic limit. To this aim, let us
first consider the case without contact interaction, {\it i.e.}, $\alpha_0\to 0$.
In this case, owing to translation invariance, we immediately get

\begin{equation}
\label{pf}
Z_0(\beta)=\int_0^\infty dE\ \rho(E)e^{-\beta E}=
\pi\left(\frac{2mE_g}{h^2}\right)^{3/2}\frac{I(\beta E_g)}{\beta E_g}\; .
\end{equation}

\noindent The corresponding density $\rho(E)$ of the one-particle quantum states, {\it i.e.}, the number of the one-particle quantum states {\it per}
unit volume and within the energy interval $E$ and $E+dE$, $E\ge 0$,
can be obtained as the inverse Laplace transform of the above expression (\ref{pf}).
Explicit evaluation yields

\begin{equation}
\label{dos0}
\rho(E)=\pi\sqrt{E_g}\left(\frac{2m}{h^2}\right)^{3/2}
\left[\frac{E}{E_g}Ai^2\left(-\frac{E}{E_g}\right)+
{Ai^\prime}^2\left(-\frac{E}{E_g}\right)\right]\ .
\end{equation}

\noindent As a consequence, we can write the average density of particles in the form

\begin{equation}
\label{dens}\frac{\left<N\right>_T}{V}\equiv
\left<n\right>_{T}=z
\int_{0}^{\infty}dE\
\frac{\rho(E) e^{-\beta E}}{1-ze^{-{\beta} E}}
+\frac{z}{V(1-z)}
\end{equation}
\noindent where $z=e^{{\beta}\mu}$.
\noindent From the above expression it immediately follows that there is no Bose-Einstein condensation in the presence of a uniform
gravitational field, owing to the behavior

\begin{equation}
\label{den0}
\rho(E=0)=\sqrt{E_g}\left(\frac{2m}{h^2}\right)^{3/2}\frac{3^{-2/3}\pi}{[\Gamma(1/3)]^2}\ .
\end{equation}

\noindent Now, it is quite evident that we can rewrite eq.~(\ref{dens}) in the
suggestive form

\begin{eqnarray}
\nonumber \left<n\right>_{T}&=&
\frac{3^{-2/3}\kappa}{2\pi[\Gamma(1/3)]^2 }\frac{g_1 (z)}{\lambda_T^2}+
z\int_{0}^{\infty}dE\
\frac{\rho(E)-\rho(0)}{1-ze^{-\beta E}} e^{-\beta E}\\
&+&\frac{z}{V(1-z)}\ ,
\end{eqnarray}

\noindent which shows that the lack of Bose-Einstein condensation in the presence of a uniform field
in three spatial dimension just corresponds to the very same phenomenon for an ideal gas of free particles in two spatial dimensions.
A little though readily drives to gather that also in the case $\alpha_0\not= 0$ the same conclusion holds true, namely condensation does not occur,
in contrast with the claim of ref.~\cite{sba}
As a matter of fact, if we set

\begin{eqnarray}
\label{fine}
\nonumber G(\alpha_0,\gamma;r)&\equiv& G(\alpha_0=0,\gamma;r)+\Delta G(\alpha_0,\gamma;r)\\
&=&\frac{I(\gamma)}{8\pi^2\gamma}+\Delta G(\alpha_0,\gamma;r)\ ,
\end{eqnarray}

\noindent it can be proved \cite{gms} that integration with respect to the radial dimensionless coordinates leads to
the finite result

\begin{equation}
\int_0^\infty dr\ r\Delta G(\alpha_0,\gamma;r)=\int_0^\infty dt
\frac{(\alpha_0\gamma)^t}{\Gamma(t+1)}\ .
\end{equation}

\noindent It immediately follows that in the thermodynamic limit the only relevant term is the first one
in the RHS of eq.~(\ref{fine}), all the rest disappearing in that limit, {\it i.e.} no Bose-Einstein condensation
is possible in the presence of a uniform field and of a point-like impurity. It should be noticed that, from the very same
results of ref.~\cite{gms}, the absence of Bose-Einstein condensation persists even if we describe the impurity
by means of an Aharonov-Bohm vortex \cite{AB}.

\end{document}